\documentclass[conference, onecolumn]{IEEEtran}

\usepackage[usenames,dvipsnames]{xcolor}
\usepackage{dsfont}
\usepackage{amsbsy}
\usepackage{amssymb, commath}
\usepackage{amsmath, mathtools, amsmath}%
\usepackage{amsthm}
\usepackage{breqn}
\usepackage{graphicx}
\usepackage{mathrsfs}
\usepackage{epstopdf}
\usepackage[font=small,labelsep=space]{caption}
\DeclareCaptionLabelSeparator{dot}{.~}
\usepackage{subcaption}
\usepackage[square,numbers, sort&compress]{natbib}
\usepackage{comment}
\usepackage{diagbox}

\newtheorem{thm}{Theorem}
\newtheorem{lem}{Lemma}

\makeatletter
\newtheorem*{rep@theorem}{\rep@title}
\newcommand{\newreptheorem}[2]{%
\newenvironment{rep#1}[1]{%
 \def\rep@title{#2 \ref{##1}}%
 \begin{rep@theorem}}%
 {\end{rep@theorem}}}
\makeatother
\newreptheorem{conj}{Conjecture}

\theoremstyle{definition}

\newtheorem{defn}{Definition}

\newtheorem{remark}{Remark}

\newcommand{\mathbbX}{\mathbb{X}}
\newcommand{\mathbbD}{\mathbb{D}}

\newcommand{\calR}{\mathcal{R}}

\newcommand{\calB}{\mathcal{B}}

\newcommand{\calM}{\mathcal{M}}

\newcommand{\calF}{\mathcal{F}}

\newcommand{\calP}{\mathcal{P}}

\newcommand{\conv}{\mathsf{conv}}
\newcommand{\sE}{\mathsf{E}}

\renewcommand{\tilde}{\widetilde}

\newcommand{\Reals}{\mathbb{R}}

\newcommand{\eps}{\varepsilon}

\newcommand{\calN}{\mathcal{N}}

\newcommand{\sEs}{\sE_{e^\eps}}

\usepackage[colorlinks=true]{hyperref}
\hypersetup{%
,urlcolor=black
,citecolor=red
,linkcolor=blue
}

\newcommand{\eqn}[2]{\begin{equation}
\label{#1}
#2
\end{equation}}
\newcommand{\al}[1]{\begin{align*}
#1
\end{align*}}

\definecolor{light-gray}{gray}{.90}

\makeatletter
\newcommand*{\addFileDependency}[1]{
  \typeout{(#1)}
  \@addtofilelist{#1}
  \IfFileExists{#1}{}{\typeout{No file #1.}}
}
\makeatother

\begin{document}

	\title{\vspace{5.5mm} A Better Bound Gives a Hundred Rounds: Enhanced Privacy Guarantees via $f$-Divergences}

\author{%
Shahab Asoodeh${}^\dagger$, Jiachun Liao${}^*$, Flavio P. Calmon${}^\dagger$, Oliver Kosut${}^*$, Lalitha Sankar${}^*$
\\
\small ${}^\dagger$Harvard University, \{shahab, flavio\}@seas.harvard.edu \\
$^*$Arizona State University, \{jiachun.liao, okosut, lalithasankar\}@asu.edu
}	
	\date{}
	\maketitle


\begin{abstract}
    We derive the optimal  differential privacy (DP) parameters of a mechanism that satisfies a given level of R\'enyi differential privacy (RDP). Our result is based on the joint range of two $f$-divergences that underlie the approximate and the R\'enyi variations of differential privacy. We apply our result to the moments accountant framework for characterizing privacy guarantees of stochastic gradient descent. When compared to the state-of-the-art, our bounds may lead to about 100 more stochastic gradient descent iterations for training deep learning models for the same privacy budget. 
\end{abstract}

\section{Introduction}

Differential privacy (DP) \cite{Dwork_Calibration} has become the \emph{de facto} standard for privacy-preserving data analytics. Intuitively, a (potentially randomized) algorithm is said to be \emph{differentially private} if its output does not vary significantly with small perturbations of the input.  DP guarantees are usually cast in terms of properties of the \emph{information density} \cite{Pinsker1964InformationAI} of the output of the algorithm conditioned on a given input---referred to as the \emph{privacy loss variable} in the DP literature. 

Several methods have recently been proposed to ensure differentially private training of machine learning (ML) models \cite{Abadi_MomentAccountant, Shokri:DeepLearning, chaudhuri2011differentially, Bassily1, Balle:Subsampling, SGD_Exponential_VS_Gaussian}. Here, the parameters of the model determined by a learning algorithm (e.g.,  weights of a neural network or coefficients of a regression) are sought to be differentially private with respect to the data used for fitting the model (i.e. the  \emph{training} data).  When the model parameters are computed by applying stochastic gradient descent (SGD) to minimize  a given  loss function, DP can be ensured by directly adding noise to the gradient. 
The empirical and theoretical flexibility of this  noise-adding procedure for ensuring DP  was demonstrated, for example, in \cite{Shokri:DeepLearning, Abadi_MomentAccountant}. This method is currently being used for privacy-preserving training of large-scale ML models in industry, see e.g., the implementation of \cite{McMahan_Privacy} in the Google's open-source TensorFlow Privacy
framework \cite{TensorFlowPrivacy}.

Not surprisingly,  for a fixed training dataset, privacy deteriorates with each SGD iteration. In practice, the DP  constraints are set \emph{a priori}, and then mapped to a permissible number of  SGD  iterations for fitting the model parameters. Thus, a key question is: \emph{given a DP constraint, how many  iterations are allowed before the SGD algorithm is no longer private?} 
The main challenge in determining the DP guarantees provided by  noise-added  SGD is keeping track of the evolution of the privacy loss random variable during subsequent gradient descent iterations.  %
This can be done, for example, by invoking  advanced composition theorems for DP, such as  \cite{Vadhan_Composition, Kairouz_Composition}. Such composition results, while theoretically significant,  may be difficult to apply to the SGD setting due to their generality (e.g., they do not take into account the noise distribution used by the privacy mechanism). 

Recently, Abadi et al.\ \cite{Abadi_MomentAccountant} circumvented the use of DP composition results by developing a method called \textit{moments accountant} (MA). Instead of dealing with DP directly, the MA approach provides privacy guarantees in terms of \emph{R\'enyi differential privacy} (RDP) \cite{RenyiDP} for which  composition has a simple linear form. Once the privacy guarantees of the SGD execution are determined in terms of RDP, they are mapped back to DP guarantees via a conversion result between DP and RDP \cite[Theorem 2]{Abadi_MomentAccountant}. This approach renders tighter DP guarantees than those obtained from  advanced composition theorems (see \cite[Figure 2]{Abadi_MomentAccountant}).

\textit{Our Contributions:} We provide a framework which settles the \textit{optimal} conversion from RDP to DP, and thus further enhances the privacy guarantee obtained  by the MA approach.   
Our technique relies on the information-theoretic study of joint range of $f$-divergences: we first describe both DP and RDP using two certain types of the $f$-divergences, namely $\sE_\lambda$ and $\chi^\alpha$ divergences (see Section~\ref{Sec:Preliminary}). We then apply \cite[Theorem 8]{Harremoes_JOintRange} to characterize the joint range of these two $f$-divergences which, in turn, leads to the ``optimal'' conversion from RDP to DP (see Section~\ref{Sec:Conversion}).  
Specifically, this optimal conversion allows us to derive bounds on the number of SGD iterations for a given DP constraint in the context of Gaussian perturbation of  the gradient. Our result improves upon the state-of-the-art \cite{Abadi_MomentAccountant} by allowing more training iterations (often hundreds more) for the same privacy budget, and thus providing higher utility for free (see Section~\ref{Sec:Gaussian}).

\section{Preliminaries and Problem Setup}\label{Sec:Preliminary}
In this section, we give several  definitions and basic results that will be used in the subsequent sections. 

Let $\mathbbD$ be some universe of all possible datasets and $(\mathbbX, \calF)$  be a measurable space with Borel $\sigma$-algebra $\calF$. A mechanism $\calM: \mathbbD\to \calP(\mathbbX)$ assigns a probability distribution $\calM_d$ to each dataset $d$ where $\calP(\mathbbX)$ denotes the set of all probability measures on $\mathbbX$. Two datasets $d$ and $d'$ are said to be neighboring (denoted by $d\sim d'$) if their Hamming distance is one. For any pair of neighboring datasets $d$ and $d'$,  the privacy loss random variable is defined as $L_{d,d'}\coloneqq \log\frac{\calM_d(Y)}{\calM_{d'}(Y)}$ where $Y\sim \calM_d$.  

\begin{defn}\label{Def_RDP}
A mechanism $\calM:\mathbbD\to \calP(\mathbbX)$ is said to be
\begin{itemize}
     \item $(\eps, \delta)$-DP for a given $\delta\in [0,1)$ 
     if 
    \eqn{DP}{\sup_{A\in \calF, d\sim d'}\calM_{d}(A)-e^\eps\calM_{d'}(A)\leq \delta.}
    \item $(\alpha, \gamma)$-RDP for a given $\alpha>1$
    , if 
    \eqn{RDP}{\sup_{d\sim d'}D_\alpha(\calM_{d}\|\calM_{d'})\leq \gamma,}
where $D_\alpha(P\|Q)\coloneqq \frac{1}{\alpha-1}\log\mathbb E_Q\left[\left(\frac{\text{d}P}{\text{d}Q}\right)^\alpha\right]$ denotes the R\'enyi divergence of order $\alpha$ between $P$ and $Q$ in $\calP(\mathbbX)$.
\end{itemize}
\end{defn}

It can be shown that \eqref{DP} is implied if the tail event $\{L_{d,d'}>\eps\}$ occurs with probability at most $\delta$ for all $d\sim d'$,  and \eqref{RDP} is implied if (and only if) the $\alpha$-moment of $L_{d,d'}$ is upper bounded by $\gamma$. Built on this intuition, the MA restricts the $\alpha$-moment of $L_{d,d'}$ for \textit{all} $\alpha>1$. 

As mentioned earlier, RDP (and hence MA) composes linearly, as opposed to the strong composition theorem for DP which is known to be loose for many practical mechanisms, including Gaussian.  With this clear advantage comes a shortcoming: RDP suffers from the lack of operational interpretation, see e.g., \cite{Balle2019HypothesisTI}. 
To address this issue, the RDP guarantee is often translated into a DP guarantee via the following result. 

\begin{thm}(\cite[Thm 2]{Abadi_MomentAccountant}, \cite[Prop 3]{RenyiDP})
\label{Theorem:RDP_Delta}
If the mechanism $\calM$ is $(\alpha, \gamma)$-RDP, then it satisfies $(\eps, \delta)$-DP for any $\eps>\gamma$ and 
\begin{equation}\label{Eq:Delta1}
    \delta = e^{-(\alpha-1)(\eps- \gamma)}.
\end{equation}
\end{thm} 
For MA, this constraint must hold for \textit{all} $\alpha>1$ and thus it leads to $(\eps, \delta)$-DP for
\begin{equation}\label{Eq:MA_DElta}
    \delta = \inf_{\alpha>1}e^{-(\alpha-1)(\eps- \gamma(\alpha))},
\end{equation}
where $\gamma (\alpha) = \sup_{d\sim d'}D_\alpha(\calM_d\|\calM_{d'})$ and the dependence on $\alpha$ is made clear.
Since $\alpha\mapsto (\alpha-1)D_{\alpha}(P\|Q)$ is convex \cite[Corollary 2]{van_Erven} for any pair of probability measures $P$ and $Q$, the above minimization is a log-convex problem and hence can be solved to an arbitrary accuracy.  We will show in Section~\ref{Sec:Gaussian} that this minimization has a simple form for Gaussian mechanisms and can be solved analytically. 

Theorem~\ref{Theorem:RDP_Delta} establishes a relationship for converting RDP to DP that is extensively used in several recent differentially private ML applications, e.g., \cite{Balle2019mixing, Balle:Subsampling, RDP_SemiSupervised, RDP2_PosteriorSampling, Feldman2018PrivacyAB, Balle_AnalyticMomentAccountant, bhowmick2018protection} to name a few.
However, despite its extensive use, this relationship is loose. For instance, as we see later, for Gaussian mechanisms this relationship holds for $\eps\to 0$ only when the variance of noise goes to infinity. In Section~\ref{Sec:Conversion}, we present the \textit{optimal} conversion from RDP to DP, thus improving the privacy guarantees of recent ML applications involving MA. Specifically, we investigate the following two closely-related questions: 

 \textbf{Question One:} \textit{Given an $(\alpha, \gamma)$-RDP mechanism $\calM$, what are the smallest $\eps$ and $\delta$ such that $\calM$ is $(\eps, \delta)$-DP?}

\noindent We show  in Section~\ref{Sec:Conversion} that such minimal $\eps$ and $\delta$ can be obtained via a simple one-variable optimization problem.  
 


We then turn our attention to privacy guarantees in applications where the data may need to be accessed many times (say $T$ times) such as with SGD. In such applications, each data access renders the application of a privacy mechanism, i.e., $T$ privacy mechanisms are applied. An oft-used model, that we also adopt here, is one in which each mechanism adds Gaussian noise with pre-specified variance $\sigma^2$. This model is referred to as the $T$-fold homogeneous
composition of Gaussian mechanisms each with variance $\sigma^2$.   
 
 \textbf{Question Two:} \textit{
 Given $\eps\geq 0$, $\delta\in [0,1]$ and $\sigma^2$,  
 what is the largest $T$ such that the $T$-fold homogeneous composition of  Gaussian mechanism with variance $\sigma^2$ is $(\eps,\delta)$-DP?}
 
The linearity of the RDP guarantee (in $T$) and the optimal conversion from RDP to DP (addressed in Question One) enable us to express the answer to this question as a minimization (over $\alpha>1$) of the answer to 
Question One, analogous to \eqref{Eq:MA_DElta}.  Although this additional minimization significantly complicates the analytic derivation, we  nevertheless obtain tight bounds for the largest $T$ provided that $\delta$ is sufficiently small. Details are deferred to Section~\ref{Sec:Gaussian}.  

To mathematically formulate these goals, we need the following definitions and basic results.

\begin{defn}(\cite{Csiszar67, Ali1966AGC})\label{eq:fdivergence}
	Given two probability distributions $P$ and $Q$ and a real-valued convex function $f$ satisfying $f(1)=0$, the $f$-divergence  between $P$ and $Q$ is given by 
	\begin{align}
		D_f(P\|Q) \coloneqq \mathbb \mathbb E_Q\left[f\left(\frac{\text{d}P}{\text{d}Q}\right)\right].
	\end{align}
	\end{defn}
	We frequently use two particular instances of $f$-divergences.   Given $\lambda\geq 1$, the $f$-divergence associated with $f(t) = (t-\lambda)_+ =\max\{t-\lambda, 0\}$, is called $\sE_{\lambda}$-divergence (also known as  \textit{hockey-stick} divergence \cite{hockey_stick}) and given by  
	\begin{equation}
	\label{Defi_HS_Divergence}
		\sE_{\lambda}(P\|Q) = \int (\text{d}P-\lambda \text{d}Q)_+=\sup_{A\in \calF} \left[P(A) - \lambda Q(A)\right].
	\end{equation}
Also, for any $\alpha>1$, the $f$-divergence associated with $f(t)=\frac{1}{\alpha-1}(t^\alpha-1)$ is denoted by\footnote{$\chi^\alpha$-divergence is also referred to as $\alpha$-Hellinger divergence, see, e.g., \cite{fdivergence_Sason16}.} $\chi^\alpha(P\|Q)$. Note that $D_\alpha(P\|Q) = \frac{1}{\alpha-1}\log\left(1 + (\alpha-1)\chi^\alpha(P\|Q)\right)$ for a pair of probability distributions $P$ and $Q$. 

It is shown in \cite{Barthe:2013_Beyond_DP}, \cite{Improving_Gaussian} that 
\begin{equation}\label{DP_HS}
    \calM~\text{is}~(\eps, \delta)\text{-DP} \Longleftrightarrow \sup_{d\sim d'}\sE_{e^\eps}(\calM_d\|\calM_{d'})\leq \delta.
\end{equation}
Similarly, it can be verified that:
\begin{equation}\label{RDP-chi-alpha}
    \calM~\text{is}~(\alpha, \gamma)\text{-RDP} \Longleftrightarrow\sup_{d\sim d'}\chi^\alpha(\calM_d\|\calM_{d'})\leq \chi(\gamma),
\end{equation}
where 
\begin{equation}\label{tilde_Gamma}
   \chi(\gamma) \coloneqq \frac{e^{(\alpha-1)\gamma}-1}{\alpha-1}.
\end{equation}


For any $\alpha>1$ and non-negative 
$\gamma$, we let $\mathbb M_\alpha(\gamma)$ be the set of all $(\alpha, \gamma)$-RDP mechanisms $\calM$. This definition, together with \eqref{DP_HS}, enables us to precisely formulate Question One. If a mechanism is $(\alpha, \gamma)$-RDP then the smallest $\delta$, for a given $\eps$, such that it is $(\eps, \delta)$-DP is upper bounded by 
\begin{equation}\label{Fundamen_Smallest_Delta}
    \delta_\alpha^\eps(\gamma) \coloneqq \sup_{\calM\in \mathbb M_\alpha(\gamma)}~ \sup_{d\sim d'}~\sE_{e^\eps}(\calM_d\|\calM_{d'})
\end{equation}
Given $\alpha, \gamma$ and $\eps$, this quantity corresponds to the smallest $\delta$ guaranteed by the worst mechanism in $\mathbb M_\alpha(\gamma)$, thus establishing an upper bound for the smallest $\delta$ such that a given $(\alpha, \gamma)$-RDP mechanism is $(\eps, \delta)$-DP. In fact, we can write 
\begin{equation}
    \delta_\alpha^\eps(\gamma) =  \inf\left\{\delta\in [0,1]
    :\forall \calM\in \mathbb M_\alpha(\gamma)~\text{is}~ (\eps, \delta)\text{-DP}\right\}. 
\end{equation}
Such quantity is key for indicating the ``optimality'' of a  conversion from RDP to DP. It may be equivalently identified by closely related quantities
\begin{equation}\label{FundamentalGamma}
        \gamma_\alpha^\eps(\delta)\coloneqq \sup\left\{\gamma\geq 0 :\forall \calM\in \mathbb M_\alpha(\gamma)~\text{is}~ (\eps, \delta)\text{-DP}\right\},
\end{equation}
or 
\begin{equation}\label{FundamentalEps}
    \eps_\alpha^\delta(\gamma)\coloneqq \inf\left\{\eps\geq 0:\forall \calM\in \mathbb M_\alpha(\gamma)~\text{is}~ (\eps, \delta)\text{-DP}\right\}.
\end{equation}
In the next section, we exploit \eqref{DP_HS}--\eqref{RDP-chi-alpha} to compute or bound these quantities. 



\section{Optimal Conversion from RDP to DP}
\label{Sec:Conversion}
In this section, we aim at computing the fundamental worst-case DP privacy parameter guaranteed by an $(\alpha, \gamma)$-RDP mechanism; a quantity defined in  \eqref{Fundamen_Smallest_Delta}. To this goal, we first show that this quantity is an upper boundary of a convex set defined by $\sE_{\lambda}$-divergence and $\chi^\alpha$-divergence and then invoke the well-known result of \cite{Harremoes_JOintRange} about the joint range of $f$-divergences. 

First note that, according to \eqref{RDP-chi-alpha},  the set $\mathbb M_\alpha(\gamma)$ can be equivalently characterized by the constraint $\chi^\alpha(\calM_d\|\calM_{d'})\leq \chi(\gamma)$, where $\chi(\gamma)$ is defined in \eqref{tilde_Gamma}. Hence, the quantity in \eqref{Fundamen_Smallest_Delta} in fact constitutes the upper boundary of the convex set 
\eqn{ConvexSet1}{\calR_\alpha\coloneqq \left\{\left(\chi^\alpha(\calM_{d}\|\calM_{d'}), \sEs(\calM_{d}\|\calM_{d'})\right)\Big|\forall\calM, d\sim d'\right\}.}This simple observation has two key implications. First, the convexity of this set implies that the map $\gamma\mapsto \delta_\alpha^\eps(\gamma)$, defined  in \eqref{Fundamen_Smallest_Delta}, can be alternatively expressed by $\delta\mapsto \gamma_\alpha^\eps(\delta)$, defined in \eqref{FundamentalGamma}. Note also that $\gamma_\alpha^\eps$ can be  equivalently written  as
\begin{align}
	\label{eq:Optimization_Gamma_DP}
	 \gamma_\alpha^\eps(\delta)= &\inf_{\calM:\mathbbD\to \calP(\mathbbX)}\inf_{d\sim d'} \chi^{-1}(\chi^\alpha(\calM_{d}\|\calM_{d'}))\\
	&\qquad \text{s.t.~} \sEs(\calM_{d}\|\calM_{d'})\geq \delta,~ \forall d\sim d'\nonumber,
\end{align}
where $\chi^{-1}(\cdot)$ is the inverse of $\chi(\cdot)$, defined in \eqref{tilde_Gamma}, and in given by 
$\chi^{-1}(t) =  \frac{1}{\alpha-1}\log(1+(\alpha-1)t)$.  
Second, to derive the upper boundary of $\calR_\alpha$ (and thus $\gamma_\alpha^\eps(\delta)$) it suffices to characterize $\calR_\alpha$. This  allows us to cast the problem of converting from $(\alpha, \gamma)$-RDP to $(\eps, \delta)$-DP as characterizing the joint range of $\sE_\lambda$ and $\chi^\alpha$ divergences. To tackle the latter problem, we refer to \cite{Harremoes_JOintRange} whose main result is as follows. 
 \begin{thm}(\cite[Theorem 8]{Harremoes_JOintRange})\label{thm:f-divergence_joint-region}
	We have  
	\begin{equation*}
		\Big\{\big(D_{f}(P\|Q), D_{g}(P\|Q)\big)\Big | P, Q\in\calP(\mathbbX) \Big\}		= {\sf conv}(\calB)
	\end{equation*}
where $\conv(\cdot)$ denotes the convex hull operator and  
$$\calB\coloneqq \Big\{\big(D_{f}(P_{\sf b}\|Q_{\sf b}), D_{g}(P_{\sf b}\|Q_{\sf b})\big)\Big | P_{\sf b}, Q_{\sf b}\in\calP(\{0, 1\}) \Big\}.$$
 \end{thm}
This theorem provides an efficient method for characterizing the joint range of any pair of $f$-divergences. Specialized to $\chi^\alpha$ and $\sE_\lambda$ divergences, this theorem therefore enables us to characterize $\calR_\alpha$ and thus derive $\gamma_\alpha^\eps(\delta)$. 
We formalize this intuition in Theorem~\ref{thm:optimization_formulation_gamma} and establish a simple variational formula for $\gamma_\alpha^\eps(\delta)$ involving a one-parameter log-convex minimization.   
Hence,  the optimization \eqref{eq:Optimization_Gamma_DP}, which can potentially be of significant complexity, turns into a simple tractable problem. 
\begin{thm}\label{thm:optimization_formulation_gamma}
For any $\alpha>1$, $\eps\geq 0$ and $\delta\in [0,1)$,  
\begin{equation}
\label{eq:Optimization_gamma}
    \gamma^\eps_\alpha(\delta)= \eps + \min_{p\in(\delta, 1)}\frac{1}{\alpha-1}\log \left( p^\alpha (p-\delta)^{1-\alpha}+\bar{p}^\alpha(e^\eps-p+\delta)^{1-\alpha}\right),
\end{equation}
	where $\bar p \coloneqq 1-p$.
\end{thm}
It can be shown the term inside the logarithm is convex in $p$ and hence this optimization problem can be numerically solved with an arbitrary accuracy. It seems, however, not simple to analytically derive $\gamma^\eps_\alpha(\delta)$. Nevertheless, we obtain a tight lower bound in the following theorem. 
\begin{thm}\label{Thm:Lower_BOund_Gamma}
	For any $\eps\geq 0$ and $\alpha>1$, we have 
	\begin{align}
	    \gamma^\eps_\alpha(0)&=0,\nonumber\\
	    \gamma^\eps_\alpha(\delta) &= \eps-\log(1-\delta), \qquad \qquad~~~~~  \text{if}~~\alpha\delta\geq 1, \label{Eq:Gamma_LB1}\\
	    \gamma^\eps_\alpha(\delta)&\geq \max\{ g(\alpha, \eps, \delta), f(\alpha, \eps, \delta)\}, ~~\textrm{if}~~~ 0<\alpha\delta< 1, \label{Eq:Gamma_LB2}
	\end{align}
where 	
	\begin{equation*}
		g(\alpha, \eps, \delta)\coloneqq \eps - \frac{1}{\alpha-1}\log\frac{\zeta_\alpha}{\delta},
	\end{equation*}
	with $\zeta_\alpha\coloneqq \frac{1}{\alpha}\left(1-\frac{1}{\alpha}\right)^{\alpha-1}$ and 
	\begin{equation*}
		f(\alpha, \eps, \delta)\coloneqq \eps + \frac{1}{\alpha-1}\log\left(\left(e^{\eps }-\alpha  \delta \right) \left(\frac{\delta -1}{\delta -e^{\eps }}\right)^{\alpha }+\alpha  \delta\right).
	\end{equation*}
\end{thm}
\begin{figure}
    \centering
    \includegraphics[height=0.3\linewidth, width=0.45\linewidth]{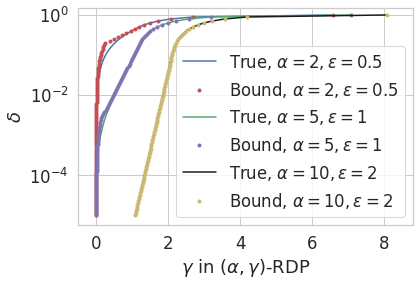}
    \caption{True values (solid curves), obtained via numerically solving convex optimization problem \eqref{eq:Optimization_gamma}, versus the bounds (dotted curves)  obtained from Theorem~\ref{Thm:Lower_BOund_Gamma} for three pairs of $(\alpha, \eps)$.}
    \label{fig:Compare_Alpha}
\end{figure}

In Fig.~\ref{fig:Compare_Alpha}, we numerically solve \eqref{eq:Optimization_gamma} for three pairs of $(\alpha, \eps)$ and compare them with their corresponding bounds obtained from Theorem~\ref{Thm:Lower_BOund_Gamma}, highlighting the tightness of the above lower bound.


As indicated earlier and illustrated in Fig.~\ref{fig:Compare_Alpha}, the lower bound in   $\gamma_\alpha^\eps(\delta)$ in Theorem~\ref{Thm:Lower_BOund_Gamma} is translated into an upper bound on $\delta_\alpha^\eps(\gamma)$. In practice, it is often more appealing to design differentially private mechanisms with a hard-coded value of $\delta$ (as opposed to the fixed $\eps$). To address this practical need, we convert the lower bound in  Theorem~\ref{Thm:Lower_BOund_Gamma} to an upper bound on $\eps_\alpha^\delta(\gamma)$.

\begin{lem}\label{Lemma_epsilon_Approximate}
	For $\alpha>1$ and  $\gamma\geq 0$, we have 
	\eqn{}{\eps^\delta_\alpha(\gamma)=\left(\gamma +\log(1-\delta)\right)_+, ~~~\textrm{if}~~~\alpha\delta\geq 1,}
	and if $0<\alpha\delta< 1$
	\begin{align}
	    \eps^\delta_\alpha(\gamma)\leq \frac{1}{\alpha-1}\min\Big\{\Big((\alpha-1)\gamma-\log\frac{\delta}{\zeta_\alpha}\Big)_+,~\log\Big(\frac{(\alpha-1)\chi(\gamma)}{\alpha\delta}+1\Big)  \Big\},\label{Eq:Approximate_epsilon}
	\end{align}
	where $\chi(\gamma)$ is defined in \eqref{tilde_Gamma}. Moreover, $\eps_\alpha^\delta(0) = 0$.
\end{lem}
The proof of this lemma is based on writing the first-order approximation for $f$ in terms of $\delta$, thereby allowing us to invert the inequality \eqref{Eq:Gamma_LB2}. Note that $g$ is a linear function of $\eps$ and hence invertible. It must be mentioned that Balle et al. \cite[Theorem 21]{Balle2019HypothesisTI} has recently proved the bound $\eps^\delta_\alpha(\gamma)\leq \gamma - \frac{1}{\alpha-1}\log\frac{\delta}{\zeta_\alpha},$ via a fundamentally different approach which is weaker than  Lemma~\ref{Lemma_epsilon_Approximate}.

\begin{remark}\label{remark_Zero_Eps}
As an important special case, this lemma demonstrates that an $(\alpha, \gamma)$-RDP mechanism provides $(0, \delta)$-DP guarantee if $1-e^{-\gamma} < \frac{1}{\alpha}$ and $\delta\in \big[\zeta_\alpha e^{(\alpha-1)\gamma},\,\frac{1}{\alpha}\big]$.  See Appendix~\ref{Appendix_Remark} for the detailed derivation and also another sufficient condition for $(0, \delta)$-DP. Notice that this is significantly stronger than what would be obtained from Theorem~\ref{Theorem:RDP_Delta}:  $\eps_\alpha^\delta(\gamma)\leq \gamma - \frac{1}{\alpha-1}\log\delta$ from which $(0, \delta)$-DP cannot be achieved. 
\end{remark}

\section{Moments Accountant and Gaussian Mechanisms}\label{Sec:Gaussian}
Moments accountant (MA) was recently proposed by Abadi et al. \cite{Abadi_MomentAccountant} as a method to bypass advanced composition theorems \cite{Vadhan_Composition,Kairouz_Composition}. Given a mechanism $\calM$, the $T$-fold adaptive homogeneous composition $\calM^{(T)}$ is a mechanism that consists of $T$ copies of $\calM$, i.e., $(\calM^1, \dots, \calM^T)$ such that the input of $\calM^i$ may depend on the outputs of $\calM^1, \dots, \calM^{i-1}$.  
Determining the privacy parameters of $\calM^{(T)}$ in terms of those of $\calM$ and $T$ is an important problem in practice and thus has been the subject of an extensive body of research, see e.g., \cite{Vadhan_Composition, Kairouz_Composition, Abadi_MomentAccountant, Balle_AnalyticMomentAccountant}. 

Advanced composition theorems \cite{Vadhan_Composition,Kairouz_Composition} are well-known results that provide the DP parameters of $\calM^{(T)}$ for general mechanisms. However, they can be loose and do not take into account the particular noise distribution under consideration (i.e., Gaussian noise). MA was shown to significantly improve updn advanced composition theorems in specific applications such as SGD. The cornerstone of MA is the linear composability of RDP:  If $\calM^1, \dots, \calM^T$ are $(\alpha, \gamma)$-RDP, then it is shown \cite[Theorem 2]{Abadi_MomentAccountant} that $\calM^{(T)}$ is $(\alpha, \gamma T)$-RDP. This result is then translated into DP privacy parameters via Theorem~\ref{Theorem:RDP_Delta}. Since the above composability and conversion hold for all $\alpha>1$, one can obtain the \textit{best} privacy parameters by optimizing over $\alpha$ according to \eqref{Eq:MA_DElta}. More precisely, $\calM^{(T)}$ is $(\eps, \delta)$-DP with  
\begin{equation}\label{eq:Delta_MA}
    \delta = \inf_{\alpha>1}e^{-(\alpha-1)(\eps-\gamma(\alpha) T)},
\end{equation} 
for a given $\eps$ or equivalently, \begin{equation}\label{eq:eps_MA}
    \eps = \inf_{\alpha>1}\gamma(\alpha)  T - \frac{1}{\alpha-1}\log\delta,
\end{equation}
for a given $\delta$, where $\gamma(\alpha) = \sup_{d\sim d'}D_\alpha(\calM_d\|\calM_{d'})$ is the RDP parameter of the constituent mechanism 
$\calM$.

For the rest of this section, we assume $\calM$ is a Gaussian mechanism and apply Theorem \ref{Thm:Lower_BOund_Gamma} and Lemma~\ref{Lemma_epsilon_Approximate} in place of \eqref{eq:Delta_MA} and \eqref{eq:eps_MA} respectively, in order to improve the DP privacy parameters obtained by MA. 
\begin{figure}
    \centering
    \includegraphics[height = 0.3\linewidth, width = 0.45\linewidth]{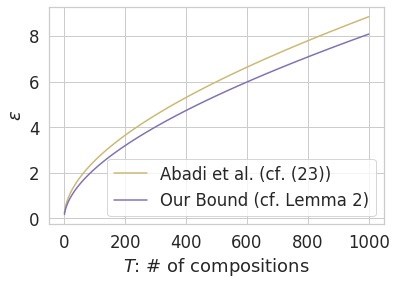}
    \caption{The comparison of our bound in Lemma~\ref{Lemma:Bound_on_Eps_MA} on  $\eps^\delta(\rho, T)$ with \eqref{eq:eps_MA2} for $\sigma = 20$ and $\delta = 10^{-5}$.}
    \label{fig:Composition}
\end{figure}
\subsection{Bounds on Privacy Parameters of Gaussian Composition}
 Let $f:\mathbbD\to \Reals^n$ be a query function and $\calM$ be a Gaussian mechanism with variance $\sigma^2$, i.e., $\mathbbX = \Reals^n$ and $\calM_d = \calN(f(d), \sigma^2\mathrm{I}_n)$ for each $d\in \mathbbD$. For simplicity, we assume that $f$ has unit $L_2$-sensitivity, i.e., $\sup_{d\sim d'}\|f(d)-f(d')\|_2=1$. Since 
 $$\sup_{d\sim d'}D_\alpha(\calM_d\|\calM_{d'}) = \frac{\alpha}{2\sigma^2}\sup_{d\sim d'}\|f(d)-f(d')\|_2 =\frac{\alpha}{2\sigma^2},$$ it follows that $\calM$ is $(\alpha, \gamma(\alpha))$-RDP  for all $\alpha>1$ where $\gamma(\alpha) = \alpha\alpha$ and $\rho=\frac{1}{2\sigma^2}$.
In light of the linear composability of RDP, we obtain that  $\calM^{(T)}$ the $T$-fold adaptive composition of $\calM$ is $(\alpha,\gamma(\alpha) T)$-RDP. In this setting, the optimization problem given in \eqref{eq:eps_MA} can be solved analytically. Consequently, MA implies that $\calM^{(T)}$ is $(\eps, \delta)$-DP for any $\delta\in (0,1)$ and
\begin{equation}\label{eq:eps_MA2}
    \eps=\inf_{\alpha>1}\gamma(\alpha) T - \frac{1}{\alpha-1}\log\delta
     = \rho T + \sqrt{4\rho T\log\frac{1}{\delta}}
\end{equation}

We next use the machinery developed in the previous section to improve \eqref{eq:eps_MA2} the DP parameter of $\calM^{(T)}$ implied by MA. To do so, define 
\begin{equation}\label{def:eps_T}
    \eps^\delta(\rho, T)\coloneqq \inf_{\alpha>1}\eps_\alpha^{\delta}(\rho\alpha T).
\end{equation}
Thus, $\calM^{(T)}$ is $(\eps^\delta(\rho, T), \delta)$-DP for any $\delta\in (0,1)$. Invoking Lemma~\ref{Lemma_epsilon_Approximate}, we can obtain a bound $\eps^\delta(\rho, T)$.  

\begin{lem}\label{Lemma:Bound_on_Eps_MA}
    The $T$-fold adaptive homogeneous composition of the Gaussian mechanism with variance $\sigma^2$ is $(\eps^\delta(\rho, T), \delta)$-DP with  $\delta\in (0,1)$ and 
    \begin{equation}
	\label{eq:epsilon_gBd}
	\eps^\delta(\rho, T)\leq \min \Big\{\eps_0(\rho, T), ~\eps_1(\rho, T),~ \Big(\frac{\rho T}{\delta} + \log(1-\delta)\Big)_+\Big\},
	\end{equation} 
    where $\rho= \frac{1}{2\sigma^2}$ and 
    \eqn{eps0}{\eps_0(\rho, T)\coloneqq \inf_{\alpha\in(1,\frac{1}{\delta}]}  \left(\rho\alpha T-\frac{1}{\alpha-1}\log\frac{\delta}{\zeta_\alpha}\right)_+,}
    \eqn{eps1}{\eps_1(\rho, T)\coloneqq  \inf_{\alpha\in(1,\frac{1}{\delta}]} \, \frac{1}{\alpha-1}\log\Big(1+\frac{e^{\rho\alpha(\alpha-1)T}-1}{\alpha\delta}\Big),}
    and $\zeta_\alpha$ is as defined in Theorem~\ref{Thm:Lower_BOund_Gamma}.
\end{lem}

\begin{figure}[t]
	\centering
 	\includegraphics[height = 0.3\linewidth, width = 0.45\linewidth]{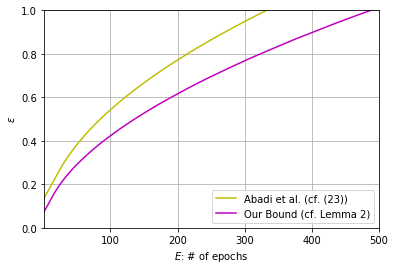}
	\caption{Privacy parameter $\eps$ of noisy SGD where the Gaussian noise with $\sigma = 4$ is added to the gradient of mini-batches with size rate (or sub-sampling rate) $0.001$. Also, $\delta$ is assumed to be $10^{-5}$.}
	\label{fig:SGD}
\end{figure}

The bound given in this lemma can shed light on the optimal variance of the Gaussian mechanism $\calM$ required to ensure that $\calM^{(T)}$ is $(\eps, \delta)$-DP (cf.\ Question Two in Section~\ref{Sec:Preliminary}). To put our result about the variance in perspective, we first mention two previously-known bounds on $\sigma^2$.
Advanced composition theorems (see, e.g., \cite[Theorem III.3]{Vadhan_Composition}) require $\sigma^2 = \Omega(\frac{T\log(1/\delta)\log(T/\delta)}{\eps^2})$. Abadi et al. \cite[Theorem 1]{Abadi_MomentAccountant} improved this result by showing that $\sigma^2$ suffices to be linear in $T$; more precisely, $\sigma^2 = \Omega(\frac{T\log(1/\delta)}{\eps^2})$. To have a better comparison with our final result, we write this result more explicitly. It follows from \eqref{eq:Delta_MA} or \eqref{eq:eps_MA} that 
\begin{align}
    \frac{T}{2\sigma^2}\leq \sup_{\alpha>1}\frac{\eps}{\alpha}+ \frac{1}{\alpha(\alpha-1)}\log\delta=\eps - 2\log\delta -2 \sqrt{(\eps-\log\delta)\log\frac{1}{\delta}},\nonumber
\end{align}
and hence assuming $\delta$ is sufficiently small, we obtain 
\begin{equation}\label{eq:variance_MA}
    \sigma^2\geq \frac{2T}{\eps^2}\log\frac{1}{\delta} + \frac{T}{\eps} + O\left(\frac{1}{\log\delta^{-1}}\right).
\end{equation}
We are now in order to state our result.
\begin{thm}\label{Theorem_noise}
The $T$-fold adaptive homogeneous composition of a Gaussian mechanism with variance $\sigma^2$ is  $(\eps, \delta)$-DP, for $\eps>2\delta\log\frac{1}{\delta}$, if 
\al{\sigma^2\geq \frac{2T}{\eps^2}\log\frac{1}{\delta} + \frac{T}{\eps} -\frac{2T}{\eps^2}\left(\log(2\log\delta^{-1})+1-\log\eps\right)+ O\left(\frac{\log^2(\log\delta^{-1})}{\log\delta^{-1}}\right).
}
\end{thm}
The proof of this theorem is based on a relaxation of Theorem~\ref{Thm:Lower_BOund_Gamma} obtained by ignoring $f(\alpha, \eps, \delta)$. Considering both $f$ and $g$ will result in a stronger result at the expense of more involved analysis. Comparing with \eqref{eq:variance_MA}, Theorem~\ref{Theorem_noise} indicates that, providing $\delta$ is sufficiently small, the variance of each constituent Gaussian mechanism  can  be reduced by $\frac{2T}{\eps^2}\left(\log(2\log\delta^{-1})+1-\log\eps\right)$ compared to what would be obtained from MA.

\subsection{Illustration of Our Bounds}\label{Sec:Abadi_SGD}
 In this section, we empirically compare our bound on $\eps^\delta(\rho, T)$ given in Lemma~\ref{Lemma:Bound_on_Eps_MA} with the privacy parameter \eqref{eq:eps_MA2} obtained via MA and has been extensively used in the state-of-the-art differentially private machine learning algorithms, e.g., \cite{Balle2019mixing, Balle:Subsampling, RDP_SemiSupervised, RDP2_PosteriorSampling, Feldman2018PrivacyAB, Balle_AnalyticMomentAccountant, bhowmick2018protection, McMahan_Privacy}. We do so in two different settings: (1) vanilla $T$-fold composition of the Gaussian mechanism with fixed variance, and (2) noisy SGD algorithm.
 
 \noindent\textbf{Vanilla Gaussian Composition:}
 Here, we wish to obtain bounds on the privacy parameter $\eps$ of $\calM^{(T)}$ where $\calM$ is a Gaussian mechanism  with $\sigma=20$. In Fig.~\ref{fig:Composition}, we compare  Lemma~\ref{Lemma:Bound_on_Eps_MA} with MA when $\delta = 10^{-5}$. According to this plot, our result enables us to achieve a smaller privacy parameter by up to $0.75$, i.e., $\max_{T\in [1000]} \eps^\delta_{\mathsf{MA}}(\rho, T)- \eps^\delta(\rho, T) = 0.75$ where $\eps^\delta_{\mathsf{MA}}(\rho, t)$ is the $\eps$ given in \eqref{eq:eps_MA2}. This privacy amplification may have important impacts on recent private deep leaning algorithms. Alternatively, one can observe that our result allows for more iteration for the same $\eps$, for instance 100 more iterations for any $\eps$ larger than $6$.  
 
 \noindent\textbf{Noisy SGD:} SGD is the standard algorithm for training many machine learning models. In order to fit a model without compromising privacy, a standard practice is to add Gaussian noise to the gradient of each mini-batch, see e.g., \cite{Abadi_MomentAccountant, Shokri:DeepLearning, Feldman2018PrivacyAB, Balle2019mixing,chaudhuri2011differentially, Bassily1, BAssily2}. The prime use of  MA  was to exploit the RDP's simple composition property in deriving the privacy parameters of the  noisy SGD algorithm \cite[Algorithm 1]{Abadi_MomentAccountant}. To have a fair comparison, we implement this algorithm with the sub-sampling rate $q=0.001$ and noise parameter $\sigma = 4$ and then compute its DP parameter via \eqref{eq:eps_MA2} with $\rho=q^2/((1-q)\sigma^2)$ (see \cite[Lemma 3]{Abadi_MomentAccountant}) and $\delta = 10^{-5}$. We then compare it in Fig.\ \ref{fig:SGD} with Lemma~\ref{Lemma:Bound_on_Eps_MA} with the same $\rho$ and $\sigma$. As demonstrated in this figure, our result allows remarkably more epochs (often over a hundred) within the same privacy budget and thus providing higher utility.


\small{
\bibliography{reference}
\bibliographystyle{IEEEtran} }
 \appendix
 \normalsize
 \section{}\label{App:Poofs}
 \subsection{Proof of Theorem \ref{thm:optimization_formulation_gamma}}
 First notice that, in light of Theorem \ref{thm:f-divergence_joint-region}, the convex set $\calR_\alpha$ defined in \eqref{ConvexSet1} is equal to the convex hull of the set $\calB_{\alpha,\eps}$ given by
 \begin{align}
 \label{eq:convex_set_RenHock_Binary}
   \calB_{\alpha,\eps}=\{(\chi^\alpha(P_{\sf b}\|Q_{\sf b}),\sEs(P_{\sf b}\|Q_{\sf b}))\big|P_{\sf b},Q_{\sf b}\in \calP(\{0,1\})\} 
 \end{align}
where $P_{\sf b}=\mathsf{Bernoulli}(p)$ and $Q_{\sf b}=\mathsf{Bernoulli}(q)$ with parameters $p,q\in(0,1)$. 
For any pair of such distributions, define $\tilde{\gamma}\coloneqq \chi^\alpha(P_{\sf b}\|Q_{\sf b}) $ and $\delta\coloneqq \sE_{e^\eps}(P_{\sf b}\|Q_{\sf b})$.  
We first show that the convex hull of $\calB_{\alpha,\eps}$ is given by 
\begin{align}
\label{eq:convex_hull_RenHock_Binary}
  \bar{\calB}_{\alpha,\eps}=\{(\tilde{\gamma}, \delta)\big| \delta\in[0,1), \tilde{\gamma}\geq \tilde{\gamma}(\delta)\} 
\end{align}
with $\tilde{\gamma}(\delta)$ given by 
\begin{align}
    \label{eq:chracterize_Binaryset}
	\tilde{\gamma}(\delta)= &\inf_{0<p,q<1}  \chi^\alpha(P_{\sf b}\|Q_{\sf b})\\
	&\qquad \text{s.t.~} \sEs(P_{\sf b}\|Q_{\sf b})\geq \delta\nonumber.
\end{align}
To this goal, we need to demonstrate that for any $\lambda\in [0,1]$ and pairs of points $(\tilde{\gamma}_1, \delta_1), (\tilde{\gamma}_2, \delta_2)\in \calB_{\alpha,\eps}$, we have  $(\lambda\tilde{\gamma}_1+\bar\lambda\tilde{\gamma}_2, \lambda\delta_1+\bar\lambda\delta_2)\in \bar{\calB}_{\alpha,\eps}$, where $\bar\lambda=1-\lambda$, or equivalently  $\lambda\delta_1+\bar\lambda\delta_2\in [0,1)$ and $\lambda\tilde{\gamma}_1+\bar\lambda\tilde{\gamma}_2\geq \tilde{\gamma}(\lambda\delta_1+\bar\lambda\delta_2)$. Hence, it suffices to show that $\delta\mapsto \tilde{\gamma}(\delta)$ is convex. 


Let $p_i,q_i\in (0,1)$ with $p_i\geq q_i$ be the optimal solution of \eqref{eq:chracterize_Binaryset} for $\delta_i$, $i=1,2$, and $P_{{\sf b},i},Q_{{\sf b},i}$ be the corresponding Bernoulli distributions. For any $\lambda\in [0,1]$, we construct two Bernoulli distribution $P_{{\sf b},\lambda}$ and $Q_{{\sf b},\lambda}$ with parameters $p_\lambda=\lambda p_1+\bar\lambda p_2$ and $q_\lambda=\lambda q_1+\bar\lambda q_2$, respectively. It can be verified that 
\begin{align}
    \sEs(P_{{\sf b},\lambda}\|Q_{{\sf b},\lambda})
    =&p_\lambda-e^\eps q_\lambda\\
    = & \lambda p_1+\bar\lambda p_2 -e^\eps(\lambda q_1+\bar\lambda q_2)\\
    \geq & \lambda \delta_1 + \bar\lambda\delta_2, 
\end{align}
i.e., $(p_\lambda, q_\lambda)$ is feasible for $\lambda\delta_1+\bar\lambda\delta_2$. In addition, from the convexity of $\chi^{\alpha}$, we have that
\begin{align}
  \lambda\tilde{\gamma}(\delta_1)+\bar\lambda\tilde{\gamma}(\delta_2) =& \lambda \chi^\alpha(P_{{\sf b},1}\|Q_{{\sf b},1})+\bar\lambda \chi^\alpha(P_{{\sf b},2}\|Q_{{\sf b},2})\\
  \geq & \chi^\alpha(P_{{\sf b},\lambda}\|Q_{{\sf b},\lambda})\\
  \geq & \tilde{\gamma}(\lambda\delta_1+\bar\lambda\delta_2).
\end{align}
Therefore, the function $\tilde{\gamma}(\delta)$ is convex in $\delta$ and hence $\bar{\calB}_{\alpha,\eps}$ is the convex hull of $\calB_{\alpha,\eps}$. In light of Theorem \ref{thm:f-divergence_joint-region}, this in turn implies that $\calR_\alpha = \bar{\calB}_{\alpha,\eps}$.


The above analysis shows that   $\delta\mapsto \tilde{\gamma}(\delta)$ in fact constitutes the upper boundary of $\calB_{\alpha,\eps}$ and thus $\calR_\alpha$. 
Since $\chi(\cdot)$ is a bijection, this allows us to deduce  
\begin{align}
    \label{eq:Optimization_Gamma_DP_Binary}
	 \gamma_\alpha^\eps(\delta) =&\inf_{0<p,q<1}  \chi^{-1}\left(\chi^\alpha(P_{\sf b}\|Q_{\sf b})\right)\\
	&\qquad \text{s.t.~} \sEs(P_{\sf b}\|Q_{\sf b})\geq \delta\nonumber,
\end{align}
and hence the optimization problem 
\eqref{eq:Optimization_Gamma_DP}  can be converted to the above two-parameter optimization problem.

Expanding both $\chi^\alpha$ and $\sE_{e^\eps}$, we can explicitly write \eqref{eq:Optimization_Gamma_DP_Binary} as
\begin{align}
\label{eq:Optimization_Gamma_binary} 
\gamma^\eps_\alpha(\delta)= &\inf_{0< q< p< 1}~ \frac{1}{\alpha-1}\log \left( p^\alpha q^{1-\alpha}+(1-p)^\alpha(1-q)^{1-\alpha}\right)\\
&\qquad \text{s.t.~} p-qe^\eps\geq\delta,\nonumber
\end{align}
where $\delta<1$ and $\gamma<\infty$. Let $h(p,q;\alpha)$ indicate the objective function of the optimization problem in \eqref{eq:Optimization_Gamma_binary}. For any given $\alpha>1$ and $ p\in (0,1)$, the partial derivative of $h(p,q;\alpha)$ with respect to $q$ is given by 
	\begin{align}
		\frac{\partial \, h(p,q;\alpha)}{\partial q} = \frac{  p^\alpha q^{-\alpha}-(1-p)^\alpha(1-q)^{-\alpha}}{ p^\alpha q^{1-\alpha}+(1-p)^\alpha(1-q)^{1-\alpha}},
	\end{align}
which is negative for all $0<q<p<1$, and therefore, $h(p,q;\alpha)$ is decreasing in $q$. In addition, for $\eps\geq 0$ and $\delta\in [0, 1)$, the two constraints $0<q<p<1$ and	$p-qe^\eps \geq \delta$ in \eqref{eq:Optimization_Gamma_binary}
can be equivalently rewritten as 
\begin{align}
	\begin{cases}
	 \delta<p<1\\
	 0<q<\frac{p-\delta}{e^\eps}.
	\end{cases}
\end{align}
Thus, the infimum in \eqref{eq:Optimization_Gamma_binary} is attained at $q=\frac{p-\delta}{e^\eps}$, and therefore, for $\alpha>1$, $\delta\in [0, 1)$ and $\eps\geq 0$, the optimization problem in \eqref{eq:Optimization_Gamma_binary} is simplified as
\begin{align}
	\label{eq:Optimization_Gamma_binary_simplified}
	e^{(\alpha-1)(\gamma^\eps_\alpha(\delta)-\eps)}= &\inf_{p}~  p^\alpha (p-\delta)^{1-\alpha}+(1-p)^\alpha(e^\eps-p+\delta)^{1-\alpha}\\
	&\text{s.t.~} \delta<p<1,\nonumber
\end{align}
which is the desired result.

 \subsection{Proof of Theorem \ref{Thm:Lower_BOund_Gamma}}
Recall that the optimization problem in Theorem~\ref{thm:optimization_formulation_gamma} is equivalent to \eqref{eq:Optimization_Gamma_binary_simplified}. 
Let $h_1(p;\alpha,\delta,\eps)$ indicate the objective function in \eqref{eq:Optimization_Gamma_binary_simplified}. One can verify that for $\alpha>1, \delta\in [0, 1)$ and $\eps>0$, the mapping  $p\mapsto h_1(p;\alpha,\delta,\eps)$ is convex. Therefore, the numerical result of $\gamma^\eps_\alpha(\delta)$ can be easily obtained for any given $\alpha,\delta$ and $\eps$.

To get closed-form expressions, we explore lower bounds of \eqref{eq:Optimization_Gamma_binary_simplified} as follows.\\
\textbf{Lower bound 1:} Ignoring the second term in  $h_1(p;\alpha,\delta,\eps)$, we obtain  
\begin{align}
	\label{eq:Optimization_Gamma_binary_simplified_LB1}
	e^{(\alpha-1)(\gamma^\eps_\alpha(\delta)-\eps)}\geq &\inf_{p}~  p^\alpha (p-\delta)^{1-\alpha}\\
	\qquad & \text{s.t.~} \delta<p<1.\nonumber
\end{align}
 We note that the objective function in \eqref{eq:Optimization_Gamma_binary_simplified_LB1} is convex in $p$. It can be observed via
\begin{align}
	\frac{\partial^2}{\partial p^2}
	p^\alpha (p-\delta)^{1-\alpha}= (\alpha-1)\alpha\left(p^{\frac{\alpha}{2}}(p-\delta)^{\frac{-1-\alpha}{2}}-p^{\frac{\alpha-2}{2}}(p-\delta)^{\frac{1-\alpha}{2}}\right)^2\geq 0,
\end{align}
and therefore, by setting the first derivative to be $0$,
we obtain the optimal solution for the the corresponding unconstrained problem as $p^*=\alpha\delta$. Since $\alpha>1$, it follows that the optimal solution of \eqref{eq:Optimization_Gamma_binary_simplified_LB1} is given by $p^*=\min\{\alpha\delta, 1\}$, and therefore
\begin{align}
	e^{(\alpha-1)(\gamma^\eps_\alpha(\delta)-\eps)}\geq    \left(\delta\alpha^\alpha (\alpha-1)^{1-\alpha}\right){\textbf{1}\{\alpha\delta< 1\}}+\left((1-\delta)^{1-\alpha}\right){\textbf{1}\{\alpha\delta\geq  1\}}
\end{align}
with equality holds if and only if $\alpha\delta\geq 1$, where $\textbf{1}\{\cdot\}$ denotes the indicator function. 
Thus, if $\alpha\delta\geq 1$, we have 
$$\gamma^\eps_\alpha(\delta)=\eps-\log(1-\delta), $$
and if $\alpha\delta< 1$, we have the lower bound 
\begin{align}
  \gamma^\eps_\alpha(\delta)\geq & \eps -\frac{1}{\alpha-1}\log\left(\frac{1}{\delta\alpha}\left(1-\frac{1}{\alpha}\right)^{\alpha-1}\right)\nonumber\\
  &= \eps -\frac{1}{\alpha-1}\log\frac{\zeta_\alpha}{\delta}. \label{Proof_g}
\end{align}

\noindent\textbf{Lower bound 2:} To obtain the second lower bound, we note that the function $h_1(p;\alpha,\delta,\eps)$ is convex in $\delta$. This enables us to bound $h_1(p;\alpha,\delta,\eps)$ from below by using its linear approximation at $\delta=0$. Hence we can write
\begin{align}
	h_1(p;\alpha,\delta,\eps)&\geq  h_1(p;\alpha,\delta=0,\eps) + \frac{\partial h_1(p;\alpha,\delta=0,\eps)}{\partial \delta}\delta\\
	&=  p + (\alpha-1)\delta +\left(\frac{1-p}{e^\eps-p}\right)^\alpha\left(e^\eps-p-(\alpha-1)\delta\right)
\end{align}
with equality if and only if $\delta=0$.
Therefore, we have
\begin{align}
	\label{eq:Optimization_Gamma_binary_simplified_LB2}
	e^{(\alpha-1)(\gamma^\eps_\alpha(\delta)-\eps)}\geq &\inf_{p}~  \left(1-\left(\frac{1-p}{e^\eps-p}\right)^\alpha\right)p +\left(\frac{1-p}{e^\eps-p}\right)^\alpha\left(e^\eps-(\alpha-1)\delta\right) + (\alpha-1)\delta\\
	& \text{s.t.~} \delta<p<1.\nonumber
\end{align}
Let $h_2(p;\alpha,\delta,\eps)$ indicate the objective function of \eqref{eq:Optimization_Gamma_binary_simplified_LB2}.
In the following, we prove the monotonicity of $h_2(p;\alpha,\delta,\eps)$ in $p$ for $\alpha>1$, $1> \delta\geq 0$ and $\eps\geq 0$.
Taking the first derivative of $h_2(p;\alpha,\delta,\eps)$ with respect to $p$, we have
\begin{align}
	\frac{\partial \,h_2(p;\alpha,\delta,\eps) }{\partial \, p}&=1+\left(\frac{1-p}{e^\eps-p}\right)^\alpha \left(\frac{\alpha(e^\eps-1)(p+(\alpha-1)\delta-e^\eps)}{(e^\eps-p)(1-p)}-1\right)\\
	&\eqqcolon h_3(p;\alpha,\delta,\eps)\nonumber\\
	\label{eq:Gamma_bd2_InPf-1}
	&\geq 1+\left(\frac{1-p}{e^\eps-p}\right)^\alpha
	 \left(-\frac{\alpha(e^\eps-1)}{1-p}-1\right)
	 \eqqcolon h_4(p;\alpha,\eps)\\
	 \label{eq:Gamma_bd2_InPf-2}
	 &>  h_4(p=\delta;\alpha,\eps)\\
	 &= \frac{(e^\eps-\delta)^{\alpha}-(1-\delta)^\alpha-\alpha(e^\eps-1)(1-\delta)^{\alpha-1}}{(e^\eps-\delta)^{\alpha}}\\
	 &\eqqcolon \frac{h_5(\delta,\alpha,\eps)}{(e^\eps-\delta)^{\alpha}}\nonumber \\
	 \label{eq:Gamma_bd2_InPf-3}
	 &\geq   \frac{h_5(\delta,\alpha,\eps=0)}{(e^\eps-\delta)^{\alpha}} =  0
\end{align}
where
\begin{itemize}
	\item the inequality in \eqref{eq:Gamma_bd2_InPf-1} is from the fact that the function $h_3(p;\alpha,\delta,\eps)$ is increasing in $\delta$, and therefore, for $1> \delta\geq 0$, $ h_3(p;\alpha,\delta,\eps)\geq h_3(p;\alpha,\delta=0,\eps) = h_4(p;\alpha,\eps)$ 
	\item the inequality in \eqref{eq:Gamma_bd2_InPf-2} is due to the fact that the function $h_4(p;\alpha,\eps)$ is increasing in $p$ as shown below  	
	\begin{align}
		\frac{\partial\, h_4(p;\alpha,\eps)}{\partial p}= \alpha(\alpha-1)(e^\eps-1)^2(1-p)^{\alpha-2}(e^\eps-p)^{-\alpha-1}>0
	\end{align}
and therefore, for $ 1>p>\delta$, $h_4(p;\alpha,\eps)>h_4(p=\delta;\alpha,\eps)$.
\item the inequality in \eqref{eq:Gamma_bd2_InPf-3} is from the monotonicity of the function $h_5(\delta,\alpha,\eps)$ in $\eps$. Specifically,
\begin{align}
	\frac{\partial \, h_5(\delta,\alpha,\eps)}{\partial\, \eps} =\alpha e^\eps\left((e^\eps-\delta)^{\alpha-1} -(1-\delta)^{\alpha-1}\right)\geq 0
\end{align}
and therefore, for $\eps\geq 0$, $h_5(\delta,\alpha,\eps)\geq h_5(\delta,\alpha,\eps=0)=0$.
\end{itemize}
Therefore, the objective function $h_2(p;\alpha,\delta,\eps)$ in \eqref{eq:Optimization_Gamma_binary_simplified_LB2} is increasing in $p$, and therefore, we have
\begin{align}
		e^{(\alpha-1)(\gamma^\eps_\alpha(\delta)-\eps)}&\geq  h_2(p=\delta;\alpha,\delta,\eps)\\
		&=\alpha\delta+ \left(\frac{1-\delta}{e^\eps-\delta}\right)^\alpha\left(e^\eps-\alpha\delta\right) \nonumber
\end{align}
with equality if and only if $\delta=0$. Thus, we have 
\begin{equation}\label{Proof_f}
    \gamma^\eps_\alpha(\delta)\geq \eps+\frac{1}{\alpha-1}\log \left(\alpha\delta+ \left(\frac{1-\delta}{e^\eps-\delta}\right)^\alpha\left(e^\eps-\alpha\delta\right) \right)
\end{equation}
where the equality holds if and only if $\delta=0$ which leads to $\gamma^\eps_\alpha(\delta=0)=0$. The lower bounds \eqref{Proof_g} and \eqref{Proof_f} give the desired result. 

\subsection{Proof of Lemma \ref{Lemma_epsilon_Approximate}}
From the first part of the proof of   Theorem~\ref{Thm:Lower_BOund_Gamma}, we have
    \begin{align}
		\eps^\delta_\alpha(\gamma)\begin{cases}
			\leq \big(\gamma-\frac{1}{\alpha-1}\log\frac{\delta}{\zeta_\alpha}\big)_{+}
			, & {\rm if  } \, \alpha\delta\leq 1\\
			= \big(\gamma +\log(1-\delta) \big)_{+}& {\rm otherwise}.
		\end{cases}
	\end{align}

    
    Next, we obtain a closed-form upper bound on $\eps^\delta_\alpha(\gamma)$ from the function $f(\alpha,\eps,\delta)$ in Theorem \ref{Thm:Lower_BOund_Gamma}. To do so, let $f_1(\alpha,\eps,\delta)$ be the expression inside the logarithm in $f(\alpha,\eps,\delta)$, i.e.,  $f_1(\alpha,\eps,\delta)\coloneqq \left(e^{\eps }-\alpha  \delta \right) \left(\frac{\delta -1}{\delta -e^{\eps }}\right)^{\alpha }+\alpha  \delta$. The second partial derivative of $f_1(\delta,\alpha,\eps)$ with respect to $\delta$ is given by
	\begin{align}
		\frac{\partial^2 \, f_1(\delta,\alpha,\eps)}{\partial\, \delta^2}
		=(\alpha-1) \alpha \left(e^\eps-1\right) \left(\frac{\delta-1}{\delta-e^\eps}\right)^\alpha \frac{\left(e^\eps \left(-2 \delta+e^\eps+1\right)-\alpha \delta \left(e^\eps-1\right)\right)}{(\delta-1)^2 \left(\delta-e^\eps\right)^2}.
	\end{align}
	Therefore, for $\alpha>1$, $\eps\geq 0$ and $1\geq\delta\geq 0$, the convexity of $f_1(\delta,\alpha,\eps)$ in $\delta$ is guaranteed by 
	\begin{align}
		\label{eq:Gamma_BD2App_condition}
		\delta-  \frac{e^{\eps}(e^\eps+1)}{2e^{\eps}+\alpha(e^{\eps}-1)}\leq 0 .
	\end{align}
	Let $f_2(\alpha,\eps)\coloneqq  \frac{e^{\eps}(e^\eps+1)}{2e^{\eps}+\alpha(e^{\eps}-1)} $, and therefore, if $\delta- f_2(\alpha,\eps)\leq 0$, we have
	\begin{align}
    \gamma^\eps_\alpha(\delta)
    &\geq  f(\alpha,\eps,\delta)=\eps +\frac{1}{\alpha-1}\log\left(f_1(\alpha,\eps,\delta)\right)\\
		&\geq  \eps +\frac{1}{\alpha-1}\log\left(f_1(\alpha,\eps,\delta=0)+ \frac{\partial \, f_1(\delta=0,\alpha,\eps)}{\partial\, \delta}\delta\right)\\
		&=\eps +\frac{1}{\alpha-1}\log\left(e^{-\eps(\alpha-1) }+\alpha\delta -\alpha\delta e^{-\eps(\alpha-1)}\right)\label{eq:eps_bound_inPf0},
	\end{align}
	with equality if and only if $\delta=0$. 
	In the following, we prove that $\delta \leq \frac{1}{\alpha}$ is a sufficient condition for $\delta- f_2(\alpha,\eps)\leq 0$ by showing that $f_2(\alpha,\eps)>1/\alpha$ for any $\alpha>1$. Taking the first partial derivative of  $f_2(\alpha,\eps)$ with respect to $\eps$, we have
	\begin{align}
		\frac{\partial \, f_2(\alpha,\eps)}{\partial\, \eps}&= \frac{e^\eps ((2+\alpha)e^{2\eps}-2\alpha e^{2\eps}-\alpha)}{(2e^\eps+\alpha(e^\eps-1))^2}\\
		&\begin{cases}
			\leq  0, & 1\leq e^\eps \leq \frac{\alpha+\sqrt{2\alpha(\alpha+1)}}{2+\alpha} \\
			> 0, & {\rm otherwise},
		\end{cases}
	\end{align}
	and therefore, 
	\begin{align}
		f_2(\alpha,\eps)-\frac{1}{\alpha}&\geq f_2\left(\alpha,\eps=\log \frac{\alpha+\sqrt{2\alpha(\alpha+1)}}{2+\alpha} \right)-\frac{1}{\alpha}\\
		&=  \frac{2(\alpha^2+\alpha(\sqrt{2\alpha(\alpha+1)}-1)-2)}{\alpha(2+\alpha)^2}\triangleq \frac{f_3(\alpha)}{\alpha(2+\alpha)^2}\\
		\label{eq:epsilon_bd2_InPf}
		 &>  \frac{f_3(\alpha=1)}{\alpha(2+\alpha)^2}=0
	\end{align}
	where the inequality in \eqref{eq:epsilon_bd2_InPf} follows from the fact that $f_3(\alpha)$ is monotonically increasing in $\alpha>1$ as shown below: 
	\begin{align}
		\frac{{\rm d} f_3(\alpha)}{{\rm d}\alpha}=\frac{\sqrt{2}\alpha(1+2\alpha)}{\sqrt{\alpha(1+\alpha)}}+2\sqrt{2\alpha(1+\alpha)}+4\alpha-2>0.
	\end{align}
	Therefore, from the inequality in \eqref{eq:eps_bound_inPf0}, we have that for $\delta\leq 1/\alpha$,
	\begin{align}
		\eps^\delta_\alpha(\gamma) &\leq  \frac{1}{\alpha-1}\log\left(\frac{e^{(\alpha-1)\gamma}-1}{\alpha\delta}+1\right)\nonumber\\
		&=  \frac{1}{\alpha-1}\log\left(\frac{(\alpha-1)\chi(\gamma)}{\alpha\delta}+1\right)\nonumber
	\end{align}
	and equality holds if and only if $\gamma=0$, i.e., $\eps^\delta_\alpha(\gamma=0)=0$.

\subsection{Derivation of Remark~\ref{remark_Zero_Eps} }
\label{Appendix_Remark}
Note that it can be verified that $\gamma-\frac{1}{\alpha-1}\log\frac{\delta}{\zeta_\alpha}<0$ for $\delta>\zeta_\alpha e^{(\alpha-1)\gamma}$. Combined with $\alpha\delta\leq 1$, we therefore have  $\eps_\alpha^\delta(\gamma) = 0$ for   $\delta\in[\zeta_\alpha e^{(\alpha-1)\gamma},\frac{1}{\alpha}]$. To have a valid non-empty interval, we must have 
the condition $\zeta_\alpha e^{(\alpha-1)\gamma}<\frac{1}{\alpha}$ that is simplified to $1-e^{-\gamma}\leq \frac{1}{\alpha}$. 
A similar holds for the case $\alpha\delta>1$: we have $\gamma +\log(1-\delta)<0$ if $\delta>1-e^{-\gamma}$. Hence, $\eps^\delta_\alpha(\gamma)=0$ if $\delta>\max\{1-e^{-\gamma},\frac{1}{\alpha}\}$. 
    
\subsection{Proof of Lemma \ref{Lemma:Bound_on_Eps_MA}}
Recall that for the $T$-fold composition of Gaussian mechanism with variance $\sigma^2$, we have  $\gamma(\alpha)=\alpha \rho T$ where $\rho=1/{\sigma^2}$.
From Lemma \ref{Lemma_epsilon_Approximate}, we have that for $\alpha\delta\geq 1$ and $0<\delta< 1$,
\begin{align}
    \eps_\alpha^{\delta}(\rho\alpha T)=\left(\rho\alpha T +\log(1-\delta)\right)_{+}
\end{align}
and therefore, 
\begin{align}
    \eps^\delta(\rho, T) &=\inf_{\alpha>1}\eps_\alpha^{\delta}(\rho\alpha T)\\
    &\leq \inf_{\alpha\geq \frac{1}{\delta}}\left(\rho\alpha T +\log(1-\delta)\right)_{+}\\
    &= \left(\frac{\rho T}{\delta} +\log(1-\delta)\right)_{+}.\label{eq:GaussianComp_InPf_alphadelta>1}
\end{align}
In addition, from  Lemma \ref{Lemma_epsilon_Approximate}, we have that for $0<\alpha\delta< 1$,
	\begin{align}
	    \eps^\delta_\alpha(\alpha\rho T)&\leq \min\Big\{\Big(\alpha\rho T-\frac{1}{\alpha-1}\log\frac{\delta}{\zeta_\alpha}\Big)_{+},\nonumber\\
	    &\qquad\qquad \frac{1}{\alpha-1}\log\Big(\frac{(\alpha-1)\chi(\alpha\rho T)}{\alpha\delta}+1\Big)  \Big\},
	\end{align}
where $\chi(\alpha\rho T)=\frac{e^{\rho\alpha(\alpha-1)T}-1}{\alpha-1}$, and therefore, 
\begin{align}
    \eps^\delta(\rho, T) =& \inf_{\alpha>1}\eps_\alpha^{\delta}(\rho\alpha T)\\
    \leq &\inf_{1<\alpha< \frac{1}{\delta}}\min\Big\{\Big(\alpha\rho T-\frac{1}{\alpha-1}\log\frac{\delta}{\zeta_\alpha}\Big)_+,\frac{1}{\alpha-1}\log\Big(\frac{e^{\rho\alpha(\alpha-1)T}-1}{\alpha\delta}+1\Big)  \Big\}\label{eq:GaussianComp_InPf_alphadelta<1}.
\end{align}
Combining the two inequalities in \eqref{eq:GaussianComp_InPf_alphadelta>1} and \eqref{eq:GaussianComp_InPf_alphadelta<1}, we obtain the upper bound of $\eps^\delta(\rho, T)$ in Lemma \ref{Lemma:Bound_on_Eps_MA}.

\subsection{Proof of Theorem \ref{Theorem_noise}}
Lemma \ref{Lemma:Bound_on_Eps_MA} illustrates that the $T$-fold adaptive homogeneous composition of the Gaussian mechanism with variance $\sigma^2$ is $(\eps, \delta)$-DP where 
\begin{align} \label{Variance_1}
  \eps =  \inf_{1<\alpha\leq  \frac{1}{\delta}}\frac{\alpha T}{2\sigma^2}- \frac{1}{\alpha-1}\log\frac{\delta}{\zeta_\alpha}.
\end{align}
Rearrenging the above, we obtain 
\begin{align} \label{Variance_2}\sigma^2 = \inf_{1<\alpha\leq  \frac{1}{\delta}}\frac{\alpha T}{2\eps+ \frac{2}{\alpha-1}\log\frac{\delta}{\zeta_\alpha} }
\end{align}
Assuming that $\frac{2\log \delta^{-1}}{\eps}\leq \frac{1}{\delta}$, or equivalently $\eps\geq 2\delta\log \delta^{-1}$, then we can plug $\alpha = \frac{2\log \delta^{-1}}{\eps}$ in \eqref{Variance_2} to obtain 
 
\begin{align}
    & \frac{\alpha T}{2\eps+ \frac{2}{\alpha-1}\log\alpha\delta- 2\log\left(1-\frac{1}{\alpha}\right) }\bigg|_{\alpha=\frac{2\log \delta^{-1}}{\eps}}\label{eq:sigma_bd_InPf1}\\
    &\qquad = \frac{ (\eps-2 \log\frac{1}{\delta})T\log\frac{1}{\delta}}{\eps^2 \left(\eps-\log\frac{1}{\delta}+\frac{-\eps+2 \log\frac{1}{\delta}}{\eps} \log \left(\frac{-\eps+2 \log\frac{1}{\delta}}{2 \log\frac{1}{\delta}}\right)- \log \left(\frac{2 \log\frac{1}{\delta}}{\eps}\right)\right)}\label{eq:sigma_bd_InPf2}\\
    &\qquad =  \frac{2T \log\frac{1}{\delta} }{\eps^2}+\frac{T}{\eps}-\frac{2T \left( \log \left(2 \log\frac{1}{\delta}\right)+1-\log \eps\right)}{\eps^2}+\frac{T}{2\eps^2 \log\frac{1}{\delta}}\nonumber\\
    &\qquad ~~~~~\cdot\Bigg(4 \log ^2\left(\frac{\log\frac{1}{\delta^2}}{\eps}\right)-6 \eps \log \left(\frac{\log\frac{1}{\delta^2}}{\eps}\right)+8 \log \left(\frac{\log\frac{1}{\delta^2}}{\eps}\right)+2 \eps^2-5 \eps+4\Bigg)+O\left(\frac{1}{\log^2\frac{1}{\delta}}\right)\label{eq:sigma_bd_InPf3}\\
    &\qquad =  \frac{2T}{\eps^2}\log\frac{1}{\delta} + \frac{T}{\eps} -\frac{2T}{\eps^2}\left(\log(2\log\delta^{-1})+1-\log\eps\right) + O\left(\frac{\log^2(\log\delta^{-1})}{\log\delta^{-1}}\right) \label{eq:sigma_bd_InPf4}.
\end{align}
where 
\begin{itemize}
    \item the expression in \eqref{eq:sigma_bd_InPf1} is from the expression of $\zeta_\alpha= \frac{1}{\alpha}\left(1-\frac{1}{\alpha}\right)^{\alpha-1}$ (defined in Theorem \ref{Thm:Lower_BOund_Gamma})
    and the condition $\eps>2\delta\log\delta^{-1}$,
    \item the expression in \eqref{eq:sigma_bd_InPf3} is the Taylor expansion of \eqref{eq:sigma_bd_InPf2} at $\delta=0$,
    \item in \eqref{eq:sigma_bd_InPf3} as $\delta\to 0$, we have $\log \delta^{-1}\to \infty$, therefore, for any fixed finite $\eps$ and  $T$, the fourth term is of order 
    $O\left(\frac{\log^2(\log\delta^{-1})}{\log\delta^{-1}}\right)$ and dominates $O\left(\frac{1}{\log^2\delta^{-1}}\right)$. 
\end{itemize}
It is worth mentioning that the choice of $\alpha$ has already appeared in literature, see e.g., \cite[Discussion following Theorem 35]{Feldman2018PrivacyAB}.

\end{document}